\documentstyle[12pt,epsf]{article}
\begin{document}

\begin{center}
{Preprint Institute for Basic Research IBR-TP-09-05}

{September 5, 2005}

\vspace*{1.00cm}
{\bf THE NOVEL ''CONTROLLED INTERMEDIATE NUCLEAR FUSION'' AND ITS POSSIBLE INDUSTRIAL REALIZATION AS
PREDICTED BY HADRONIC MECHANICS AND CHEMISTRY}

{\bf  Ruggero Maria  Santilli}

{Institute for  Basic Research}

{P. O. Box  1577, Palm Harbor,  FL 34682,  U.S.A.}

{ibr@ibr.net, http://www.i-b-r.org}

\end{center}

\begin{abstract}
In this note, we propose, apparently for the first time, a new type of controlled nuclear fusion
called "intermediate" because occurring at energies intermediate between those of the ''cold'' and
''hot'' fusions, and propose a specific industrial realization. For this purpose: 1) We
show that known limitations of quantum
mechanics, quantum chemistry and special relativity cause excessive departures from the
conditions occurring for all controlled fusions; 2) We outline the
covering  hadronic mechanics, hadronic chemistry and isorelativity specifically conceived,
constructed and verified during the past two decades for new cleans energies and fuels; 3) We
identify seven physical laws predicted by the latter disciplines that have to  be verified by all
controlled nuclear fusions to occur; 4) We review the industrial
research conducted to date in the selection of the  most promising engineering realization as well as
optimization of said seven laws; and 5) We propose with construction details a  specific {\it
hadronic reactor}  (patented and international patents pending), consisting of  actual equipment
specifically intended for the possible industrial production of the clean energy released by
representative  cases of controlled intermediate fusions for independent scrutiny by interested
colleagues.
\end{abstract}

\vskip0.30cm

\noindent {\bf 1. Limitations of Quantum Mechanics, Quantum Chemistry and Special Relativity.}
Following the pioneering research by Fleishmann, Pons and Hawkins [1a] of 1989 vast
research [1b,1c,1d] has confirmed the existence
of Low Energy Controlled Nuclear Fusions (LECNF) popularly called ''Cold Fusion'' (CF). Nevertheless,
the achievement of an industrially relevant energy output has remained evasive, and none is in sight
at the moment on strict scientific grounds.

Similarly, additional research supported by a collective international investment
of about one billion  dollars has shown that the High Energy Controlled Nuclear Fusion (HECNF),
popularly called ''Hot Fusion'' (HF), can indeed be attained in laboratory, although this research
too has not achieved results of  industrial significance, and none is in sight at the moment, due to
uncontrollable instabilities at the initiation of the fusions and other reasons.

In view of the above protracted insufficiencies at  high and low energies, in this paper we propose,
apparently for the first time, a new type of nuclear synthesis under the name of
Intermediate Energy Controlled Nuclear Fusions (IECNF), or ''Intermediate Fusion'' (IF) for short.

A main argument is that, for the case of the ''cold fusion'', the available energy is insufficient
for a systematic exposure of nuclei via the control of electron clouds, in
which case no fusion is evidently possible. For the case of the ''hot fusion'' we have the opposite
occurrence in which the available energy is simple excessive, thus preventing the
possibility of a real control. The name ''intermediate'' is here proposed to denote that
the available energy is indeed intermediate between those of the ''cold'' and ''hot'' fusions.

More particularly, the available energy for the proposed
intermediate fusion is set to a value sufficient for the ionization of atoms in order to expose
nuclei in preparation for their controlled synthesis, as it is the case for a plasma created by an
electric arc. Such a plasma is typically at about $10,000^o F$, thus having an energy that cannot be
credibly qualified as belonging to either the ''coldÕÕ or the ''hot'' fusion.

It should be indicated that numerous plasmas have been used in the ''cold fusion'' research
[1b,1c,1d]. Nevertheless, dramatic differences will soon emerge between the ''intermediate fusion''
proposed in this paper and existing plasma fusion research due to irreconcilable differences in the
assumed basic disciplines.

Another objective of this paper is to propose specific reactors, called for technical reasons
explained below {\it hadronic reactors} (patented and international patents pending) for the possible
industrial utilization of the clean energy expected from ''intermediate fusions.'' To achieve this
task we shall: 1)  Identify the basic disciplines that are applicable to
all controlled fusions, whether ''cold,'' ''intermediate'' or ''hot''; 2) Identify the
basic laws that have to be verified for any controlled fusion to occur; and 3) Propose with
all necessary construction details a specific hadronic reactor based on the realization and
optimization of said physical laws.

To begin, let us recall that the original discovery of the cold fusion [1a] created considerable
controversies fueled by authoritative voices stating that {\it cold fusions are not predicted by
quantum mechanics.} Subsequently, the existence of CF was experimentally established. Yet, with the
passing of time, said authoritative doubts have been ignored and the CF has been essentially studied
to this day via the use of quantum mechanics.

Another objective of this paper is to resume the authoritative doubts on the incompatibility
between CF and established disciplines because the inability to achieve industrially valid results
for both the ''cold'' and ''hot'' fusions may well be due to insufficiencies of the basic
disciplines.

To begin our analysis, recall that limitations on the universal validity of quantum mechanics,
quantum, chemistry and special relativity were fully known by the middle of the past century. As an
example, this author became a theoretical physicist because of  the doubts in the ''completion'' of
quantum mechanics expressed by Einstein, Podolsky and Rosen [2a], or the doubts
expressed by Fermi (Ref. [2b], p. 111 when treating the structure of nuclei) ''as to whether the
usual concepts of geometry hold for such small region of space" and other authoritative doubts.

With the passing of the decades, debates on these authoritative doubts were
suppressed in most technical journals. However,  increasingly cataclysmic events caused by our
alarming environmental problems have rendered mandatory the search for new clean energies
and fuels. In turn, said need forced the conduction in the last part of the 20-th century of
systematic studies [3] on the {\it limitations} of quantum mechanics, quantum chemistry and special
relativity, that is, the identification of the conditions under which said disciplines can be
safely assumed as being exactly valid, and the different  conditions under
which said disciplines are only approximately valid.

The latter studies [3] (see also independent
monographs [4] and comprehensive literature covering two decades of
research in Ref. [3i]), have confirmed  that quantum mechanics, quantum chemistry and special
relativity can indeed be assumed to be {\it exactly valid} under the conditions of
their original conception, construction and verification, namely,  {\it for systems of
point-like particles and electromagnetic waves propagating in vacuum (empty space).}

Typical examples of exact validity of quantum mechanics and special relativity are
the structure of the hydrogen atom, particles moving in high energy
accelerators, the structure of crystals and various other structures for which the
indicated conditions of applicability are met.

However, studies [3,4] have identified precise conditions under
which quantum mechanics and special relativity remain evidently applicable, but are
only {\it approximately valid,} among which we note:

1) In {\bf particle physics,} there exist various cases in which the fit of
experimental data requires the introduction of arbitrary parameters, as it is the
case for the Bose-Einstein correlation that requires {\it four} arbitrary
parameters. In reality, these parameters constitute  a direct
measure of the {\it deviations} of the Bose-Einstein correlation from the
unadulterated axioms of quantum mechanics and special relativity. As an example,
the two point correlation function of the Bose-Einstein correlation requires
off-diagonal terms that are incompatible with the quantum axiom of expectation  values of
operators that, to be observables, must be Hermitean, thus diagonal (see
monographs [3i,3k] for details).

2) In {\bf atomic physics,} quantum mechanics and special relativity have not
permitted an  exact representation of all spectral data of the helium, with
embarrassing deviations from the experimental
data of heavy atoms such as the zirconium, let alone the
historical inability in about one century to understand
the spectral emission of the Sun (see, again, refs. [3i,3k] for details).

3) In {\bf nuclear physics,} quantum mechanics and special relativity have been
unable to represent the experimental data of the simplest
possible nucleus, the deuteron, because of the inability to explain the spin 1
of its ground state (since quantum axioms require that the ground state of two
particles with spin 1/2 should be 0, while the ground state of the deuteron has spin 1),
the lack of an exact representation of the deuteron magnetic moment despite all possible
relativistic corrections, the historical inability to understand the stability of the
neutron in the deuteron, and other basic insufficiencies, with truly embarrassing
deviations from experimental data of heavy nuclei [3i,3k].

4) In {\bf superconductivity,} quantum mechanics, quantum chemistry and special relativity have
created a condition similar to that of atomic physics prior to the
representation of the structure of atoms, since said disciplines cannot explain the
bond of the two identical electrons in the Cooper pair (evidently because electrons
repel each other according to quantum mechanics), thus resulting in  a description of
an {\it ensemble}
  of Cooper pairs without a true description of their
structure [3l].

5) In {\bf chemistry,} quantum mechanics, quantum chemistry and special relativity have been unable
to provide an exact representation of the binding energy of the
simplest molecule, the hydrogen molecule (due to the
historical 2\% missing when using unadulterated quantum
axioms), with larger deviations when passing to more
complex molecules such as water (for which, e.g., electric and
magnetic moments are predicted with
the wrong sign, let alone large numerical deviations), not to forget the
embarrassing prediction by quantum chemistry that all molecules are ferromagnetic (a direct
consequence of the independence of the electrons in valence bonds, thus permitting the
polarization of their orbits under an external magnetic field). At the same time,
adulterations of quantum axioms now vastly used to improve the approximation, such
as the so-called "screenings of the Coulomb law," imply the abandonment of the very
quantum of energy (because no longer admitted for  potentials of the type
${q_1q_2\over r}e^{f(r)}$), while the same screenings imply structural departures
from quantum axioms (because the transition from the Coulomb to  screened
potentials  requires {\it nonunitary} transforms, thus exiting from the classes of
equivalent of quantum mechanics) (see [3l] for details).

6) In {\bf biology,} any claims of exact validity of quantum
mechanics, quantum chemistry and special relativity constitute scientific deceptions because, as
experts are expected to know to qualify as such,
quantum treatments imply that biological structures are
perfectly rigid, perfectly reversible in time and perfectly eternal, as it is
typically the case for crystals (see monograph [3j] for details).

7) In {\bf engineering,} various equipment show sizable deviations from
quantum mechanics, quantum chemistry and special relativity. An illustration important for the
reactors to be proposed in this note, the use of
Maxwell's equations and quantum chemistry for underwater electric arcs between
graphite electrodes is afflicted by a {\it ten fold error in excess} in the prediction of
the produced carbon monoxide, a {\it ten-fold error in defect} in the production
of carbon dioxide in the combustion exhaust, a {\it fourteen-fold error in
defect} in the amount of oxygen present in the combustion exhaust, and other
deviations simply too big to be accountable via the usual {\it ad hoc} parameters or
other manipulations to adapt  reality to preferred theories.

All the above limitations exist for {\it matter.} Additional large insufficiencies
exist for {\it antimatter} as presented in details in monograph [3n], such as the inability by
Einsteinian theories to provide a consistent {\it classical} treatment of antiparticles because,
after quantization, one obtains a ''particle,'' rather than a charge conjugated ''antiparticle''
with the wrong sign of the charge, the impossibility for Einsteinian theories to
provide a distinction between classical neutral bodies made up of matter and antimatter, and
other serious insufficiencies. These insufficiencies are ignored hereon
because we shall evidently deal with nuclei made up of {\it matter.}
Nevertheless, the insufficiencies for antimatter are sufficient, alone, to establish the
nonscientific nature of any claim of terminal character of quantum mechanics, quantum chemistry and
special relativity.

Independently from these studies, a mere confrontation of  reality with the
basic axioms of quantum mechanics, quantum chemistry and special relativity is sufficient to
establish their limits of exact applicability.

As an example, a confrontation of the pillar of special relativity, the
 Poincar\'e symmetry, and the structure of hadrons, nuclei, molecules and stars
is sufficient to see the {\it impossibility} for special relativity to be exactly
valid for the structures considered.

In fact, a necessary condition for the validity of the Poincar\'e symmetry,
well known to experts to qualify as such, is to have a {\it Keplerian structure} as
occurring in the atomic and planetary structures, while {\it hadrons, nuclei,
molecules, stars and other systems do not have a Keplerian structure because, e.g., nuclei
do not have nuclei.} The modification of the Poincar\'e symmetry to account for the
absence of Keplerian nuclei, no matter how small, causes its evident breaking, with
consequential {\it impossibility} beyond scientific doubt for special relativity to be
exactly valid for the structures considered (collectively called {\it interior dynamical
systems}, while atomic and planetary systems are examples of {\it exterior dynamical
systems} [3a,3b]).

Similarly, an inspection of the basic dynamical equations of quantum mechanics and quantum chemistry
is sufficient to see the {\it impossibility} for the theories to be exactly valid in
interior dynamical systems. In  fact, said disciplines are  based on the familiar
Scr\"odinger equation
$$
i \times  {\partial\over  \partial
t}|{\psi}> =  H(t, r, p) {\times} |{\psi}>,
\eqno(1.1a)
$$
$$
 p_k {\times}|{\psi}> = -  i {\times} {\partial}_k|{\psi}>, \; \; \;   H = {p^2\over 2\times m} +
V(r),
\eqno(1.1b)
$$
and the equivalent Heisenberg equations for a (Hermitean) observable A,  expressible in
their finite and infinitesimal forms
$$
U\times U^\dag = U^\dag \times U = I
\eqno(1.2a)
$$
$$
A(t) = U(t)\times A(0)\times U^\dag(t) = e^{H\times t\times i}\times A(0)\times
e^{-i\times t\times H},
\eqno(1.2b)
$$
$$
i{\times} {d A\over dt} = [A, H] = A {\times} H  -  H {\times} A,
\eqno(1.2c)
$$
with related canonical commutation rules
$$
[r^i,  p_j] =  i\hat {\times} {\delta}^i_j,\; \; \;
 [r^i, r^j] = [p_i, p_j] = 0.
\eqno (1.3)
$$
where $\hbar = 1$ and we have used the symbol $\times$ to denote the conventional
associative product of quantum mechanics in order to distinguish it from other products
needed for this paper.

Inspection of the above equations confirms their exact validity for systems of
point-like particles moving in vacuum, but also identifies the impossibility to
represent, e.g., the hyperdense fireball of the Bose-Einstein correlation, or the deep overlapping
of electrons in valence bonds and their breaking in chemical reactions  due to the strictly linear,
local, potential and reversible character of the equations, while said fireball or chemical
reactions are expected to be dominated by nonlinear, nonlocal,  nonpotential and irreversible
effects.

Alternatively and equivalently, the impossibility for the above equations to be
exactly valid for conditions  1)-7) above can be derived from
their basic Euclidean topology, since the latter solely admit the treatment of a
finite set of isolated points.

Consequently, any claim of exact validity of Eqs. (1.1)-(1.3) for the
Bose-Einstein correlation, chemical reactions and other processes is nonscientific, since the only
scientifically debatable issue is the identification of the applicable {\it generalization} of
quantum mechanics and special relativity.

When passing to the study of controlled nuclear fusions, the {\it impossibility} of quantum
mechanics, quantum chemistry and special relativity to be exactly valid for
becomes dramatic (see [3k] for details). Besides limitations 1)-7), we here restrict
ourself to the indication that {\it quantum mechanics, quantum chemistry and special relativity are
strictly invariant under time reversal, trivially, because all known potentials are
reversible in time. Consequently, said disciplines can only predict the synthesis
of nuclei in a form that is time reversal invariant, that is, by equally admitting
as causal their disintegration. This feature alone, let alone numerous other
technical inconsistencies [3i,3k], is sufficient to establish that quantum mechanics, quantum
chemistry and special relativity are not suited for
quantitative treatment of any controlled nuclear fusion, whether cold, intermediate or hot.}

In closing this section it should be indicated that the use of the word
''violation'' of quantum mechanics, quantum chemistry and special relativity would not be
scientifically appropriate because said disciplines were not conceived and
constructed for the conditions considered (e.g., antimatter had yet to be discovered when
Einstein formulated the special and general relativities). This illustrates the reason for the use of
the word ''inapplicable.''

It should be finally indicated that, contrary to quasi-religious beliefs
during the second half of the 20-th century, the insufficiencies of quantum
mechanics, quantum chemistry and special relativity are {\it multiplied,} rather than
resolved or even decreased, under the assumption that the
hypothetical quarks are physical constituents of hadrons existing in our
spacetime.

As an illustration, the reduction of the deuteron to quarks {\it increases} the
difficulties in representing the magnetic moment, trivially, because the hypothetical orbits
of the hypothetical quarks are too small to yield polarizations sufficient to fit
experimental data, while the problem of the spin of the deuteron is equally multiplied,
trivially, because of difficulties for quark conjectures to represent the spin of
individual protons and neutrons, let alone their bond, not to mention lack of
achievement of a true confinement that bypasses the prediction from Heisenberg's
uncertainty principle of a finite probability that quarks are free in dramatic
disagreement with available experimental evidence.

Any study of controlled nuclear fusions via quark conjecture is dismissed in this paper also because
{\it quarks cannot experience gravity,} since gravity can only be defined in our
spacetime, while quarks are purely mathematical representations of a purely mathematical
internal unitary symmetry defined on a purely mathematical complex-valued unitary space
without any credible definition in our spacetime (that is prohibited by
O'rafearthaigh's theorem).

Stated in plain language, no serious studies of controlled nuclear fusions can be credibly voiced
under the assumption that  the nuclei to be fused can freely float in space due to their
reduction to hypothetical quarks without any provable gravity.

In this note, we shall assume that the SU(3)-color, Mendeleev-type
classification of hadrons into families is of final character, and we shall assume
quarks what they are technically, purely mathematical objects useful for said
classification, while we shall assume that the physical constituents of hadrons are
basically unknown at this writing.

By recalling that the historical contributions to civilization produced by
molecules, atoms and nuclei were based on the capability {\it to extract the
constituents free,} the assumption of the
hypothetical quark as physical particles permanently bound inside hadrons is
considered nowadays one of the biggest obstructions  against
new clean energies so much needed by mankind [3k,3m].
\vskip0.50cm

\noindent {\bf 2. The Covering Hadronic Mechanics, Hadronic Chemistry and Isorelativity.}
Studies [3,4] have established that the basic insufficiency responsible for
{\it all} limitations 1)-7) of the preceding section is the impossibility to
represent interactions due to deep mutual penetration and overlapping of the wavepackets
and/or charge distributions of particles as illustrated in Figure 1. We are here referring
to {\it interactions of  contact (thus zero-range) type, nonlinear (in the
wavefunctions), nonlocal-integral (because occurring in a finite
volume), and nonpotential, thus not representable with a Hamiltonian.}

This limitation is
evidently due to the fact that quantum mechanics, quantum chemistry and special relativity
are strictly linear, local-differential and potential theories. Consequently, the
interactions depicted in Figure 1 are beyond any hope of
representation.



\begin{figure}
\begin{center}
\epsfxsize=5cm
\parbox{\epsfxsize}{\epsffile{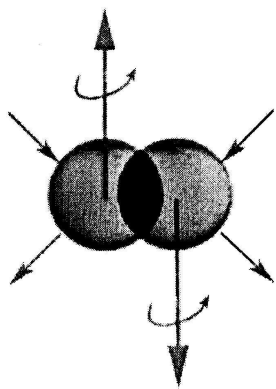}}
\end{center}
\caption{{\it A schematic view of the  interactions at the foundation of hadronic
mechanics and isorelativity, those caused by  deep wave-overlappings of the charge
distribution as well as of the wavepackets of particles. A main purpose of this paper is
to show that these interactions are crucial for industrial realizations of LENS.
}}\label{Fig1}
\end{figure}


A typical illustration is that of valence bonds that are abstracted by quantum
chemistry into two point-particles interacting at a distance. It is, of course,
true that electrons have a {\it point-like charge.} However, the idea that
electrons have a ''point-like wavepacket'' is outside the boundary of serious
science. When this physical reality is admitted, valence bonds result to be due
not only to electromagnetic interactions but also to contact nonlocal-integral and
nonpotential interactions due to the mutual penetration of the wavepackets as
depicted in Figure 1.

The  lack of representation of deep wave-overlappings has been proved to
be responsible for the lack of $2\%$ of experimental data in molecular binding
energies [3l,9a,9b], the  departures from spectral data in the helium (where,
contrary to popular belief, the two electrons are partially in conditions of
mutual overlap as in Figure 1) [3k], and  other insufficiencies.

When at the Department of Mathematics of
Harvard University in the late 1970s, R. M. Santilli initiated comprehensive
research toward a solution of the insufficiencies of conventional doctrines outlined
in Section 1.

The central problem was {\it to identify a broadening-generalization of quantum
mechanics, quantum chemistry and special relativity in such a way to represent  linear,
local and potential interactions, as well as additional, contact,
nonlinear, nonlocal-integral and nonpotential interactions.}

Since the Hamiltonian can only represent conventional interactions, the
above condition requested the identification of a {\it  new quantity} capable
of representing  interactions that, by conception, are outside the capability of
a Hamiltonian. Another necessary condition was the exiting from the class of
equivalence of quantum mechanics, as a consequence of  which the broader theory
had to be {\it nonunitary,} namely, its time evolution has to violate the
unitarity condition (1.2a).  The third and most insidious condition was the {\it
invariance,} namely, the broader theories had to represent the new nonpotential
interactions in a way as invariant as that of conventional interactions, so as to
predict the same numerical values under the same conditions at different times. We
assume that experts are aware of the {\it theorems of catastrophic
inconsistencies of noncanonical and nonunitary theories} [5j], which theorems
mandate the achievement of invariance for any theory to have physical value.

It was evident that a solution verifying the above conditions required {\it
new mathematics, e.g. new numbers, new spaces, new geometries, new symmetries,
etc.} A detailed search in advanced mathematics libraries of the
cantabridgean area revealed that the need new mathematics simply did not
exist.

Following additional (unpublished) trials and errors, Santilli [5a,5b] proposed in
1978 the solution consisting in the representation
of said contact, nonlinear, nonlocal and nonpotential interactions via a {\it
generalization (called lifting) of the basic unit} $+1$ of conventional theories into a
function, a matrix or an operator $\hat I$ that is positive-definite like $+1$, but
otherwise has an arbitrary functional dependence on all needed quantities, such
as time $t$, coordinates $r$, momenta
$p$, density $\mu$, frequency $\omega$, wavefunctions $\psi$, their derivatives
$\partial \psi$, etc.
$$
+1 \; \; \; \rightarrow \; \; \; \hat I(t, r, p, \mu. \omega, \psi. \partial\psi,
...) = 1 / \hat T > 0,
\eqno(2.1)
$$
while jointly lifting the conventional associative product
$\times$ between two generic quantities $A, B$ (numbers, vector fields, matrices,
operators, etc.) into the form admitting
$\hat I$, and {\it no longer}
$+1$, as the correct left and right unit
$$
A\times B \; \; \; \rightarrow \; \; \; A\hat \times B = A\times \hat T\times B,
\eqno(2.2a)
$$
$$
1\times A = A\times 1 = A \; \; \; \rightarrow \; \; \; \hat I\hat \times A = A\hat
\times \hat I = A,
\eqno(2.2b)
$$
for all elements $A, B$ of the set considered.

The selection of the basic unit resulted to be unique for the verification of the
above three conditions. As an illustration, whether generalized or not, the unit is
the basic invariant of any theory. The representation of non-Hamiltonian
interactions with the basic unit permitted the crucial by-passing of the
theorems of catastrophic inconsistencies of nonunitary theories [5j] (skeptic
readers are encouraged to try alternative solutions).

Since the unit is the ultimate pillar of all mathematical, physical and chemical
formulations, liftings (2.1) and (2.2) requested a corresponding, compatible
lifting of the {\it totality} of the mathematical, physical and chemical formulations used by
conventional theories, resulting indeed into new numbers, new
fields, new spaces, new algebras, new geometries, new symmetries, etc, [3,4].
This explains the dimension and time
requested by the research. Following the original proposal of 1978 to
build hadronic mechanics [5a,5b], mathematical maturity in the formulation of
the new numbers was reached in memoir [5c] of 1993 and general mathematical
maturity was reached in memoir [5d] of 1996. Physical maturity was then quickly
achieved in papers [5e,5f,5g].

In honor of Einstein's vision on the lack of completion of quantum mechanics,
Santilli submitted in the original proposal [5a,5b] the name of {\it isotopies} for
the above liftings, a word used in the Greek meaning of ''preserving the original
axioms.'' In fact, $\hat I$ preserves all topological properties of $+1$, $A\hat
\times B$ is as associative as  the conventional product $A\times B$ and the
preservation of the original axioms holds at all subsequent levels to such an extent
that, in the event any original axiom is not preserved under isotopies, the
lifting is incorrect. Nowadays, the resulting new mathematics is known as {\it
Santilli isomathematics} [4], $\hat I$ is called {\it Santilli's isounit},
$A\hat \times B$ is called the {\it isoproduct,} etc.

The fundamental dynamical equations of hadronic Mechanics were submitted by Santilli
in the original proposal [5a], are today called {\it Heisenberg-Santilli
isoequations,} and can be written in the finite form
$$
\hat U\hat \times \hat U^\dag = \hat U\hat \times \hat U = \hat I \not = 1,
\eqno(2.3a)
$$
$$
\hat A(\hat t) = \hat U(\hat t)\hat \times \hat A(\hat 0)\hat \times \hat
U^\dag(\hat t) =
(\hat e^{ \hat H\hat \times  \hat t\hat \times \hat i})\hat \times \hat A(\hat 0)
\hat
\times (\hat e^{-\hat i\hat \times \hat t\hat \times \hat H}) =
$$
$$
= [(e^{H\times \hat T\times t\times i})\times \hat I]\times \hat T\times A(0) \times
\hat T\times [\hat I\times (e^{-i\times t\times \hat T \times H})] =
$$
$$
(e^{H\times \hat T\times t\times i})\times \hat A(\hat 0) \times (e^{-i\times t\times \hat
T
\times H}),
\eqno(2.3b)
$$
and infinitesimal form [5a,5g]
$$
\hat i\hat {\times} {\hat d\hat A\over \hat d\hat t} =
i\times \hat I_t\times {d\hat A\over d\hat t} = [\hat A\hat ,
\hat H] = \hat A\hat {\times}\hat H \hat - \hat H\hat {\times}\hat A =
$$
$$
= \hat A\times \hat T(\hat t, \hat r, \hat p, \hat \psi, \hat
\partial \hat \psi, ...)\times \hat H - \hat H\times \hat T(\hat t,
\hat r, \hat p, \hat \psi, \hat
\partial \hat \psi,
...)\times \hat A,
\eqno(2.4)
$$
where: Eq. (2.3a) represent the crucial {\it isounitary property,} namely, the
reconstruction of unitarity on {\it iso-Hilbert spaces} over isofields with inner
product $<\hat \psi|\hat \times |\hat \psi>$; all quantities have a ''hat'' to
denote their formulation on isospaces over isofields with isocomplex numbers $\hat c =
c\times \hat I$, $c\in C$; and one should note the {\it isodifferential calculus} with
expressions of the type $\hat d /\hat d\hat t = \hat I_t\times d / d\hat t$ first
achieved in memoir [5d] (see below).

The equivalent lifting of Schr\"odinger's equation was  suggested by Santilli
[5a,6a], Myung and Santilli [6b] and by Mignani [6c], all original proposals
being formulated on conventional spaces over conventional fields. The isoequation was
reformulated via the isodifferential calculus by Santilli [5d], it is today called the
{\it Schr\"odinger-Santilli isoequation,} and can be written
$$
\hat i\hat {\times} {\hat {\partial}\over \hat {\partial}\hat
t}|\hat {\psi}> = i \times \hat I_t\times {\partial\over \partial \hat t} |\hat \psi> =
\hat H\hat {\times} |\hat {\psi}> =
$$
$$
= \hat H(\hat t, \hat r, \hat p)\times \hat T(\hat r, \hat p, \hat
{\psi}, \hat {\partial}\hat {\psi},. ...)\times |\hat {\psi}> = \hat
E\hat {\times} |\hat {\psi}> = E\times |\hat \psi>,
\eqno(2.5)
$$
with {\it isoexpectation values}
$$
<\hat A> = {<\hat \psi|\hat \times \hat A\hat\times |\hat \psi>\over <\hat \psi|\hat
\times |\hat \psi>}
\eqno(2.6)
$$
and basic properties
$$
{<\hat \psi|\hat \times \hat I\hat\times |\hat \psi>\over <\hat \psi|\hat \times
|\hat \psi>}
 = \hat I,\; \; \hat I\hat {\times} |\hat {\psi}> = |\hat {\psi}>,
\eqno(2.7a)
$$
$$
\hat I^{\hat n} = \hat I\hat \times \hat I\hat \times ... \hat I \equiv \hat I, \; \; \;
\hat I^{\hat {1/2}} = \hat I,
\eqno (2.7b)
$$
confirming that $\hat I$ is indeed the isounit of hadronic mechanics (where the
isoquotient $\hat / = /\times \hat I$ has been tacitly used [5d]).

By the mid 1990's, despite the isotopic lifting of  all possible quantities
and operations, hadronic mechanics had not yet reached an invariant formulation. In
particular, hadronic mechanics still missed a consistent representation of the
isotopic momentum, thus preventing systematic verifications and applications.

 After extensive
additional studies, the problem resulted to rest where nobody would expect it, in
the ordinary differential calculus. It was popularly believed for centuries in
mathematics that the differential calculus is independent from the unit of
the underlying field. Such a belief is, of course, correct, for {\it constant
units.} However, isomathematics uses isounits with an arbitrary functional
dependence that does require the lifting into the {\it isodifferential
calculus}, that permitted the first invariant
formulation of the {\it isomomentum} [4d]
$$
 \hat p_k\hat {\times}|\hat
{\psi}> = - \hat i\hat {\times}\hat {\partial}_k|\hat {\psi}> = -
i\times \hat I_k^i\times \partial_i|\hat {\psi}>,
\eqno(2.8)
$$
with {\it isocanonical commutation rules}
$$
[\hat r^i\hat ,  \hat p_j] = \hat i\hat {\times}\hat {\delta}^i_j =
i\times \delta^i_j\times \hat I, [\hat r^i, \hat r^j] = [\hat p_i,
\hat p_j] = 0.
\eqno (2.9)
$$

A few comments are now in order. Note the identity of Hermiticity and its
isotopic image, $(<\hat \psi|\hat \times \hat H^{\hat \dag})\hat \times |\hat \psi> \equiv
<\hat \psi|\hat \times (\hat H\hat \times |\hat \psi>),  \hat H^{\hat \dag} \equiv \hat
H^\dag$, thus implying that all quantities that are observable for quantum mechanics
remain observable for hadronic mechanics; the new  mechanics is indeed isounitary, thus
avoiding the theorems of catastrophic inconsistencies of nonunitary theories; hadronic
mechanics preserves all conventional quantum laws, such as Heisenberg's uncertainty principle,
Pauli's exclusion principle, etc.; hadronic mechanics has been
proved to be ''directly universal'' for all
possible theories with conserved energy, that is, capable of representing all infinitely
possible systems of the class admitted (universality) directly in the frame of the observer without
transformations (direct universality); and numerous other features one can study in
Refs. [3i,3k,3l].

Note the crucial {\it representation of irreversibility under the conservation of
the total energy,} as necessary for  isolated irreversible processes such as controlled nuclear
fusions,
$$
\hat T(t, ...) = \hat T^\dag(t, ...) \not = T(-t, ...), \; \; \; i\times \hat d\hat
H/\hat d\hat t = [\hat H\hat ,\hat H] \equiv 0.
\eqno(2.10)
$$

Also, one should note that {\it hadronic mechanics verifies the abstract axioms
of quantum mechanics to such an extent that the two mechanics coincide at the abstract,
realization-free level.} In reality, hadronic mechanics provides an explicit and
concrete realization of the theory of ''hidden variables'' $\lambda$ [2c], as one can
see from the abstract identity of the isoeigenvalue equation $H\hat \times |\hat
\psi> = \hat E\hat \times |\hat \psi>$ and the conventional  equation
$H\times |\psi> = E\times |\psi>$, by providing in this way an {\it operator}
realization of hidden variables $\lambda = \hat T$ (for detailed studies on these aspects,
including the {\it inapplicability} of Bell's inequality [2d] for hadronic mechanics
due to its nonunitary structure, we refer the reader to memoir [7h]).

We should also indicate that the birth of hadronic mechanics can be seen in the following
{\it new isosymmetry,} here expressed for a constant $K$ for simplicity,
$$
<\psi|\times |\psi>\times 1 \equiv <\psi|\times K^{-1}\times |\psi>\times (K\times
1) = <\psi|\hat \times |\psi>\times \hat I.
\eqno(2.11)
$$

The reader should not be surprised that the above  isosymmetry
remained unknown throughout the 20-th century, because its identification required
the prior discovery of {\it new numbers,} Santilli's isonumbers with
arbitrary units [5c].

Compatibility between hadronic and quantum mechanics is reached via the condition
$$
Lim_{r>>10^{-13 cm}} \hat I \equiv I,
\eqno(2.12)
$$
under which hadronic mechanics recovers quantum mechanics uniquely and identically
at all levels.

The name of "hadronic mechanics" was suggested by Santilli [5a] to
represent strong interactions as well as all possible short range interactions. The
new mechanics was then constructed in such a way to coincide everywhere with
quantum mechanics except inside the so-called {\it hadronic horizon,} namely, a
sphere of radius $1 F = 10^{-13} cm$.

A simple method has been identified in Refs. [5f,5g] for the
construction of hadronic mechanics and all its underlying new mathematics. This
method is important for controlled nuclear fusions   because it permit the implementation of
existing conventional models  into covering isomodels, thus
permitting the addition of contact nonpotential interactions that will soon
acquire a crucial role for  controlled nuclear fusions. The method consists in:

(i) Representing all conventional interactions with a Hamiltonian
$H$ and all non-Hamiltonian interactions and effects with the
isounit $\hat I$;

(ii) Identifying the latter interactions with a nonunitary transform
$$
U\times U^{\dagger} = \hat I \not = I
\eqno(2.13)
$$
\noindent and

(iii) Subjecting the {\it totality} of conventional mathematical, physical
and chemical quantities and all their operations to the above nonunitary
transform, resulting in expressions of the type
$$
I\rightarrow \hat I = U\times I\times U^{\dagger} = 1/\hat T,
\eqno(2.14a)
$$
$$
a\rightarrow \hat a = U\times a\times U^{\dagger} = a\times \hat I,
\eqno(2.14b)
$$
$$
a\times b\rightarrow U\times (a\times b)\times U^{\dagger} =
$$
$$
= (U\times a\times U^{\dagger})\times (U\times
U^{\dagger})^{-1}\times (U\times b\times U^{\dagger}) = \hat a\hat
{\times}\hat b,
\eqno(2.14c)
$$
$$
e^A\rightarrow  U\times e^A\times U^{\dagger} = \hat I\times e^{\hat
T\times \hat A} = (e^{\hat A\times \hat T})\times \hat I,
\eqno(2.14d)
$$
$$[X_i, X_j]\rightarrow U\times [X_i
 X_j]\times U^\dagger =
$$
$$
=  [\hat X_i\hat ,\hat X_j] = U\times (C_{oj}^k\times X_k)\times
U^{\dagger} = \hat C_{ij}^k\hat {\times}\hat X_k =
$$
$$
= C_{ij}^k\times \hat X_k,
\eqno(2.14e)
$$
$$
<\psi | \times |\psi >\rightarrow U\times <\psi | \times |\psi
>\times U^{\dagger} =
$$
$$
= <\psi | \times U^{\dagger}\times (U\times U^{\dagger})^{-1}\times
U\times |\psi >\times (U\times U^{\dagger}) =
$$
$$
= <\hat \psi |\hat {\times} |\hat \psi >\times \hat I,
\eqno(2.14f)
$$
$$
H\times |\psi>\rightarrow U\times (H\times |\psi>) = (U\times
H\times U^{\dagger})\times (U\times U^{\dagger})^{-1}\times (U\times
|\psi>) =
$$
$$
= \hat H\hat {\times} |\hat {\psi}>, etc.
\eqno (2.14g)
$$

Note that the above simple rules permit the explicit construction of the new
isoeigenvalues equations and related iso-Hilbert space over isonumbers, as well as of all
 needed aspects, including isoalgebras, isosymmetries and their isorepresentations [3].

Note also that {\it catastrophic inconsistencies emerge in the event even one single
quantity or operation is not subjected to isotopies.} In the absence of
comprehensive liftings, we would have a situation equivalent to the elaboration of
quantum spectral data of the hydrogen atom with isomathematics, resulting of dramatic
deviations from reality.

It is easy to see that the application of an additional nonunitary
transform $W\times W^\dag \not = I$ to expressions (2.14) causes the {\it lack of
invariance,} with consequential activation of the catastrophic
inconsistencies of theorems [5j]. However, any given nonunitary transform can be identically
rewritten in the isounitary form,
$$
W\times W^{\dagger} = \hat I,\; \; \;  W = \hat W\times \hat T^{1/2},
\eqno(2.15a)
$$
$$
W\times W^{\dagger} = \hat W\hat {\times}\hat W^{\dagger} = \hat
W^{\dagger}\hat {\times}\hat W = \hat I, \eqno(2.15b)
$$
 under which hadronic mechanics is indeed isoinvariant
$$
\hat I\rightarrow \hat I' = \hat W\hat {\times}\hat I\hat
{\times}\hat W^{\dagger} = \hat I,
\eqno(2.16a)
$$
$$\hat A\hat {\times}\hat B\rightarrow \hat W\hat {\times}
(\hat A\hat {\times}\hat B)\hat {\times}\hat W^{\dagger} =
$$
$$
= (\hat W\times \hat T\times A\times \hat T\times \hat
W^\dagger)\times (\hat T\times \hat W^\dagger)^{-1}\times \hat
T\times (\hat W\times
$$
$$
\times
 \hat T)^{-1}\times (\hat W\times \hat
T\times \hat B\times \hat T\times \hat W^\dagger) =
$$
$$
= \hat A'\times (\hat W^{\dagger}\times \hat T\times \hat
W)^{-1}\times \hat B' = \hat A'\times \hat T\times \hat B' =  \hat
A'\hat {\times}\hat B',\; etc.
\eqno(2.16b)
$$
Note that the invariance is ensured by the {\it numerically
invariant values of the isounit and of the isotopic
element under nonunitary-isounitary transforms,}
$$
\hat I \rightarrow \hat I' \equiv \hat I,\; \; \;  A\hat \times B\rightarrow A'\hat \times' b'
\equiv A'\hat \times B',
\eqno(2.17)
$$
in a way fully equivalent to the invariance of quantum mechanics,
as expected to be necessarily the case due to the preservation of the abstract axioms
under isotopies. The resolution of the catastrophic inconsistencies for noninvariant
theories is then consequential.

Hadronic  mechanics has nowadays clear
experimental verifications in particle physics, nuclear physics, superconductivity,
chemistry, astrophysics, cosmology and biology (see monographs [3j,3k,3l] for details),
which verifications cannot possibly be reviewed here. We merely mention for subsequent
need for controlled nuclear fusions  the reformulation of  valence bonds via hadronic chemistry
characterized by the isounit [9a,9b]
$$
\hat I = Diag. (n_{11}^2,n_{12}^2, n_{13}^2,n_{14}^2)\times Diag. 9n_{21}^2,
n_{22}^2, n_{23}^2, n_{24}^2)\times
$$
$$
\times e^{N\times (\hat\psi / \psi)\times \int d^3r \times
\psi^\dag_{\downarrow}(r)\times \psi_{\uparrow}(r)}
\eqno(2.18)
$$
where $n_{ak}^2, a = 1, 2, k = 1, 2, 3$ are the semiaxes of the ellipsoids
characterizing the two particles, $n_{a4}, a = 1, 2$ represents their density,
$\hat \psi$ represents the isowavefunctions, $\psi$ represents the conventional
function, and $N$ is a positive constant.

The use f the above isounit permitted R. M. Santilli and D. Shillady [9a,9b] to reach
the first exact and invariant  representation on scientific records
 of all characteristics of the hydrogen, water and other molecules, said
representation being achieved directly from first axiomatic principles without any {\it
ad hoc} parameters, or screening adulterations of the Coulomb law. In reality, due to its
nonunitary structure, hadronic chemistry contains as a particular cases all infinitely
possible screenings of the Coulomb laws (see [3l] for details).

Note the admission of quantum chemistry for the atomic structure in molecular bonds
and the use of a covering chemistry only in the short range valence interactions, namely, inside the
''hadronic horizon.'' In fact, at distances sufficiently greater than $1 F$, the volume integral in the
exponent of Eq. (2.18) is identically null, the actual dimensions and density
of the particles are ignorable, and Santilli's isounit (2.18) verifies the crucial
condition (2.12).

We should also mention that, when the Schr\"odinger-Santilli isoequation is worked out in
detail under isounit (2.18), there is the emergence of a strongly attractive Hulten
potential that, as well known, behaves at short distances like the Coulomb potential, thus  absorbing the
repulsive Coulomb force between the valence electrons [9a,9b].
$$
U\times[\left(\frac{1}{2\mu_1}p_1\times p_1+\frac{1}{2\mu_2}p_2
\times p_2 + \frac{e^2}{r_{12}}-\frac{e^2}{r_{1a}}-
             \frac{e^2}{r_{2a}}-
           \frac{e^2}{r_{1b}}-\frac{e^2}{r_{2b}}+
           \frac{e^2}{R} \right)\times |\psi \rangle] \approx
$$
$$
        \approx \left(-\frac{\hbar^2}{2\times\mu_1}\times\nabla^2_1-
            \frac{\hbar^2}{2\times\mu_2}\times\nabla^2_2-
            V\times\frac{e^{-r_{12}\times b}}{1-e^{-r_{12}\times b}}-\right.
        \ \ \ \ \ \ \ \ \ \ \ \ \ \ \ \
$$
$$
        \ \ \ \ \ \ \ \ \ \ \ \ \ \ \ \
      \left.- \frac{e^2}{r_{1a}}-\frac{e^2}{r_{2a}}-
            \frac{e^2}{r_{1b}}-\frac{e^2}{r_{2b}}+
            \frac{e^2}{R}\right)\times |\hat\psi \rangle,
\eqno{(2.19)}
$$
 (see monograph [3l], chapters 4 and 5 for details). The insufficiency of quantum chemistry is now
transparent because, without the hadronic lifting, the total electromagnetic force between the
two hydrogen atoms is identically null.

Refs. [9a,9b] achieved in this way the first model of
valence bonds in scientific records with an {\it explicitly  computed and strongly attractive force
between the electrons of a valence bond,}for which reason the model is today often referred to as
the {\it Santilli-Shillady strong valence bond,} were the word "strong" evidently refers to the
strength of the valence force.

Superficial inspections of  hadronic mechanics and chemistry may tend to dismiss the
relevance of the strong valence bond for controlled nuclear fusions. Recall that the conventional
quantum view on valence are basically insufficient to explain how two electrons can
bond to each other to form the  molecular structures of our everyday life, while
having identical charges that cause extreme repulsions at the distances of valence
bonds. For this reason, the various valence models of quantum chemistry are
mere nomenclatures, because none of them identifies the {\it attractive} character of the
valence force in an {\it explicit} form, let alone with an a {\it numerical value}
sufficient to represent reality.

A crucial feature established by hadronic chemistry is that {\it the new contact,
nonlinear, nonlocal and nonpotential interactions due to wave-over\-lapping are
strongly repulsive at short distance for triplet couplings (parallel spins) and
strongly attractive in singlet coupling (antiparallel spins}.

After (and only after) the above scientific journey, the importance
of hadronic mechanics and chemistry begins to emerge for the objective that motivated their
construction, the prediction and quantitative treatment of basically {\it new}
clean energies and fuels, that is, energies and fuels NOT predicted by quantum mechanics
and chemistry.

The {\it isotopic (axiom-preserving) lifting of special relativity} required a parallel
extensive research that cannot possibly be review here. We merely mention that this
lifting, today known as {\it  isorelativity,} was first reached by Santilli as follows:
the first isotopies of the Minkowski space were presented in Ref. [7a,7b]; the first
isotopies of the rotational symmetry were reached in Ref. [7c]; the first isotopies of
the SU(2)-spin symmetry were formulated in Ref. [7d]; the first isotopies of the
Poincar\'e symmetry and special relativity were submitted in Refs. [7e]; the first
isotopies of the spinorial covering of the Poincar\'e symmetry were reached in Refs.
[7f,7g]; the first implications for local realism was reached in ref. [7h]; and the first
comprehensive studies on the iso-Minkowskian geometry was presented in Ref. [7i] (for numerous
related works, see monographs [3]).

Note that, due to the positive-definiteness of the isounit and rule (2.14c), all
isosymmetries are locally isomorphic to the original symmetries, as necessary under
isotopies, yet they provide the most general known nonlinear, nonlocal and
non-Hamiltonian realizations of known spacetime and internal; symmetries. Intriguingly,
these isosymmetries generally reconstruct as exact on isospaces over isofield all
symmetries  believed to be broken [3h,3i].

The reader should know that isorelativity is based on a {\it geometric unification of
the Minkowskian and Riemannian geometries} [7i], {\it with consequential unification of
special and general relativities} that are now differentiated by the selected
realization of Santilli's isounit. These unifications permitted a novel formulation
of gravity that is invariant under the Poincar\'e-Santilli isosymmetry. These advances have
permitted the first and perhaps only known axiomatically consistent grand
unification of electroweak and gravitational interactions [8], where the axiomatic
consistency is achieved thanks to the reformulation of gravity via the axioms of
electroweak interactions.

 The reader should also be aware that
{\it isorelativity provides the ultimate formulation of the possible industrial
realization of controlled nuclear fusions proposed in this paper.} As an illustration, certain key
features of  controlled nuclear fusions predicted by isorelativity are
dependent on the abandonment of the philosophical abstraction of the ''universal
constancy of the speed of light'' and the assumption instead that light is a local
variable $C = c/n(t, r, p, \mu, \omega, ...)$ depending on the characteristics of
the medium in which it propagates, assuming that light can propagate at all in a given
medium.

Another belief that has to be abandoned for the formulation of "new" energies  is that
the speed of light is the maximal causal speed, and the replacement with a new {\it
maximal causal speed characterized by the geometry of the medium,} that happens to be $c$
in vacuum. An illustration is given
electrons propagating in water at a speed 1/3-rd greater than the local
speed of light (Cerenkov effect), with consequential catastrophic inconsistencies in
case  special relativity is assumed to be valid within physical media.

If the speed of light {\it in vacuum} is assumed as the maximal causal speed
{\it in water,} we have the violation of the principle of causality, while if we assume
the speed of light {\it in water} to be the maximal causal speed {\it in water}, we have the
violation of the relativistic addition of speeds. The statement that special relativity is
recovered by reducing light to photons scattering among atoms has been proved to be
nonscientific because: 1) The reduction to photons of electromagnetic waves with
one meter wavelengths traveling in water with speed ${2\over 3}c$ has no credibility; 2)
The reduction to photons for light traveling faster than that in vacuum according to vast
experimental evidence now available is nonscientific; and 3) The nonscientific
character of the reduction is established by the fact that the reduction of light to photons, even
when applicable, is afflicted by an error in defect of about $30\%$, namely, it can only represent
a few percentage of the reduction of the speed of light, and not its $33\%$ reduction (due to the
very low cross section of Compton scattering as serious scientist are expected to know).

In reality, there is no need for calculations, but only to observe and admit evidence
visible to our naked eye. A source of light submerged within pure water shows no
dispersion. This implies that photons have to scatter along a straight
line to represent the lack of dispersion, a first impossibility, while on the other
side, the speed of light is decreased of about $33\%$ compared to the speed in vacuum. The
impossibility of a credible manipulation in an attempt to salvage special relativity
under these conditions is beyond credible doubt.

In conclusion, as clearly stated by Albert Einstein in his limpid writing, and as
reviewed in Section 1, special relativity was conceived, constructed and verified {\it in
vacuum}. The validity of special relativity for all conditions existing in the universe
has been proffered by Einstein's followers. Evidence beyond any
possible doubt establish that special relativity is inapplicable
for interior dynamical problems, including dynamics within physical media,  systems without a
Keplerian nucleus (as it is the case for nuclear fusions) and others.

Rather than being topics of esoterica academic interest, the above issues have direct
societal implications for the much needed new clean energies and fuels. In fact, {\it
nuclei constitute some of the densest media measured by mankind until now.} It then
follows that {\it nuclear fusions cannot be reduced to events in vacuum.}
Consequently, any insistence without clear evidence on the exact validity of special
relativity for  nuclear fusions was tolerated in the past as an act of scientific fervor, but
nowadays the potential severe injury to society forces the denunciation of such a
fervor  particularly when proffered by experts.

\vskip0.50cm


\noindent {\bf 3. Physical Laws of Cold, Intermediate and Hot Fusions as
Predicted by Hadronic Mechanics, Hadronic Chemistry and Isorelativity.} One of the first
contributions of hadronic mechanics, hadronic chemistry and isorelativity to controlled fusions is
the identification of seven different physical laws that have to be obeyed by all Controlled Nuclear
Fusions (CNF) to occur, and have to be optimized in engineering realizations for CNF to acquire
industrial relevance. These laws were first derived in ref. [3k], they apply for cold, intermediate
and hot fusions,  and are referred to
in the literature as {\it Santilli's laws for Controlled  Nuclear Fusions.} We are not in a position
to review here their derivation to avoid a prohibitive length. Nevertheless, for completeness of this
presentation we provide below their outline with a few comments.
\vskip0.30cm

{\bf LAW I: CNF must verify the conservation of the energy.} This is the trivial law that needs no
comment.
\vskip0.30cm

{\bf LAW II: The most probable CNF are those occurring under the conservation of the angular
momentum.} The differences between quantum and hadronic mechanics begin to emerge.
Conservation laws of linear momentum and angular momentum are necessary for Keplerian structures,
such as planetary or atomic systems, in which no collision among the constituents is admitted and the
constituents are assumed to be point-like. The same laws are not necessarily verified for the
broader interior systems that include collisions of extended constituents. To do serious science we
must admit that {\it during actual collisions of extended particles (such as billiard balls), linear
momentum can be transformed into angular momentum, and vice versa.} The same feature must
be kept under quantization to avoid evident inconsistencies. Needless to
say, whenever linear momentum and angular momentum transforms into each others, the sum of their
energies is conserved. Stated differently, the only conservation laws out of ten characterized the
Poincar\'e symmetry that are necessarily verified in the physical reality are the conservation of
the energy and the uniform motion of the center of mass for isolated systems. It is at this point
where isorelativity becomes mandatory to conduct serious scientific studies of CNF. In fact,  the
Poincar\'e-Santilli isosymmetry  does indeed permit the exchange of linear momentum into angular
momentum and vice-versa (under the conservation of the total energy) because occurring under the
lifting of the conventional symmetry firstly, to represent extended particles and, secondly, to
represent ''contact'' interactions as in Eq. (2.12). Exchanges of linear and angular momenta under
collisions are then consequential. Note that ''cold fusions'' may not admit energies sufficient for
the transformation of linear into angular momentum. However, these energies are definitely available
for the ''hot fusions,'' while the case of ''intermediate fusions'' requires specific studies.

\vskip0.30cm

{\bf LAW III: CNF only occur for nuclei with compatible spins given by the ''planar singlet
coupling'' or the ''axial triplet coupling'' of Figure 3.} This is another
law with profound engineering implications indicated in Section 5. This
law also illustrates the structural differences between quantum and hadronic mechanics,
as well as the necessity of the latter for CNF. The constituents of a bound state of two quantum
particles must necessarily be point-like to avoid structural
inconsistencies beginning with the local-differential topology.
Consequently, singlet and triplet couplings are equally possible for quantum mechanics. When the
actual extended character of the constituents is taken into account, it is easy to see that
{\it planar triplet couplings of extended particles at short distances are strongly repulsive,
while planar singlet couplings are strongly attractive,} where the word ''planar'' is intended to
indicate that the two nuclei have a common median plane (Figure 2). This law was introduced by
Santilli in the original proposal [5a] to build hadronic mechanics  via the so-called {\it gear
model.} In fact, the coupling of gears in triplet (parallel spins) causes extreme repulsion to the
point of breaking the gear teeth, while the only possible coupling of gears is in singlet
(antiparallel spins). As we shall see, the first reason why ''cold'' and ''hot'' fusions have not
achieved industrial relevance until now is the general lack of controlled  implementation of this
basic law.
\vskip0.30cm



\begin{figure}
\begin{center}
\epsfxsize=7cm
\parbox{\epsfxsize}{\epsffile{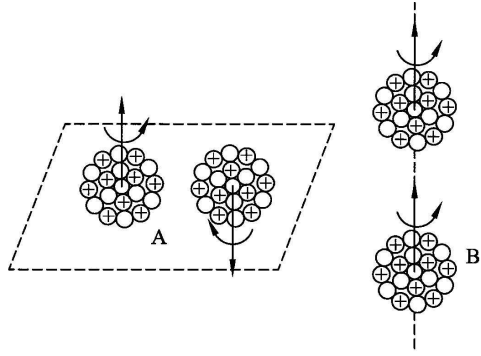}}
\end{center}
\caption{{\it A schematic view of only two stable couplings permitted by hadronic mechanics, the
''planar singlet coupling'' of the l.h.s. and the ''axial triplet coupling'' of the r.h.s. All other
spin configurations have been proved to produce strongly ''repulsive'' forces under which no CNF
is credibly possible. The configuration preferred in this paper is the axial one for reasons of
bigger efficiency in the energy output. }}\label{Fig2}
\end{figure}


{\bf LAW IV: The most probable CNF are those occurring at threshold energies (namely, at
the minimum value of the energies of the constituents needed to verify Law 1).} A main
reason of this law is that {\it all energies above the indicated threshold value cause
instability that reduce the rate of synthesis.} As we shall see, the lack of
engineering implementation of this law constitutes another reason why ''cold'' and ''hot'' fusions
have not achieved industrial relevance until now. Note  that this law favors the ''cold'' over the
''hot'' fusion. In fact, the lack of achievement of industrial significance by the ''hot fusion''
until now is particularly
due to instabilities caused by the available energies that are excessively bigger than the
indicated threshold value.
\vskip0.30cm

{\bf LAW V: The most probable CNF are those without the release of massive particles
(such as protons, neutrons and electrons).} This law was not expected by the author. Yet,
contrary to popular beliefs,  explicit calculations based on hadronic (and certainly not
quantum) mechanics indicated that the probability of a nuclear synthesis with the
release of neutrons is much {\it smaller} than that of another synthesis with the emission
of massive particles. As we shall see, this fourth law appears to be verified by nuclear syntheses
spontaneously occurring in nature.

\vskip0.30cm

{\bf LAW VI: A necessary condition for CNF to occur is to control the peripheral atomic
electrons in such a way to allow nuclei to be exposed.} Nature has set matter in
such a way that nuclei are strongly shielded by their atomic clouds. It is evident that a
''nuclear'' synthesis between two conventional ''atoms'' is impossible at low energies
because the electron clouds will never allow nuclei to approach each other, let alone
to synthesize a new nucleus. This law explains the difficulties for ''cold fusions'' to
achieve industrial significance in energy output because, by definition, ''cold fusions'' do not
have the energy necessary for the ionization of atoms. This law also illustrates the need for the
proposed ''intermediate fusions.''

\vskip0.30cm

{\bf LAW VII: CNF cannot occur without a trigger (that is, an external mechanism forcing  exposed
nuclei through the hadronic horizon).} All nuclei are positively charged, thus repelling each
other.Without a mechanism that overcomes the Coulomb repulsion and brings
nuclei inside the hadronic horizon of $1 F = 10^{-13} cm$, no nuclear synthesis is evidently
possible. However, when inside the hadronic horizon and the preceding laws are verified, the
synthesis is inevitable, as we shall see, due to the strongly attractive hadronic forces as for
model (2.19).

\vskip0.30cm

Evidently, the achievement of industrially relevant energy outputs by CNF requires
the engineering optimization of all preceding laws. This  is less obvious of what
may appear in first inspection because each law can be realized in a number of different
engineering versions. However, this does not means that all realizations have equal
efficiency. Maximization of the energy output is realized only when said engineering
realizations ''optimize'' the laws.

It is instructive to examine a representative case of ''cold fusion'' under the above physical
laws. Consider the {\it Fleishmann-Pons electrolytic cell}  [1a]. It is easy to see that
this cell does indeed verify Law 1 (conservation of the energy), Law II (conservation of the
angular momentum), Law IV (absence of excessive energy over threshold), Law V (absence of secondary
radiation) and law VII (the trigger being characterized in this case by the electrostatic pressure
compressing deuteron atoms inside the palladium).

However, Fleishmann-Pons electrolytic cell [1a] does not verify Law III (control of the
singlet  couplings) as well as Laws VI (control of atomic clouds to expose nuclei). In fact, nuclear
spin couplings occur at random, there is no clearly identified mechanism to expose nuclei, and there
is an equally clear lack of optimization of the  verified laws. Consequently, nuclear syntheses
occur at random, thus preventing  industrial values of the energy outputs.

It is an instructive exercise for researchers serious in real advances in new clean
energies to inspect other realizations of ''cold fusions'' among the large variety existing in the
literature [1b-1c]. One can see in this way that, to our best knowledge at this time, {\it none of
available ''cold fusions'' realizes ''all'' seven basic laws} (the indication of the contrary would
be appreciated).

In conclusion, Santilli's Laws on Controlled Nuclear Fusions practically  rule out the possible
achievement of industrially meaningful ''hot fusions,'' by confirming in this way a rather
widespread  consensus in the scientific community. The same laws offer serious possibilities for
''cold fusions'' to achieve industrial relevance under a number of revisions of their engineering
realizations, by therefore confirming another widespread consensus. However, the same laws identify
quite clearly the need for the proposed ''intermediate fusions'' in order to optimize their
engineering realizations.

\vskip0.50cm


\noindent {\bf 4. The New Chemical Species of Santilli's Magnecules.}
Inspection of Laws I-VII for Controlled Nuclear Fusions (CNF)  reveals that the most difficult
engineering  realization is that of Law VI on the control of electron clouds so as to expose
nuclei as a  pre-requisite for for their fusion. The author has worked for
years to achieve an industrially relevant solution of this problem (thanks to large private
investments). This section is devoted to a brief outline of the proposed solution because truly basic
for the concrete industrial realization proposed in the next section.

The current environmental problems are not caused by fossil fuels per se, but rather by
the strength of their valence bonds that has prevented the achievement of a full
combustion for over one century. In fact, hydrocarbons and other pollutants in
the exhaust literally are chunks of uncombusted molecules (for which
very reason these pollutants are carcinogenic).

A solution was proposed in Ref. [9c] of 1998
consisting of a new chemical species, today known as {\it Santilli magnecules} (in order to
distinguish them from the conventional molecules) whose bond is stable, but sufficiently
weaker than the conventional valence bond to permit full combustion (see website [9d] and
monograph [3l] for comprehensive studies).

The new species required the identification of a {\it new attractive force among
atomic constituents that is not of valence type as a central condition, thus occurring
among atoms irrespective of whether  valence electrons are available or not.}

The solution proposed in Ref. [9c] was the use of an external magnetic field sufficient
to create the polarization of atomic orbitals into toroids, as a result of which the
orbiting electrons create
a magnetic moment along the symmetry axis of the toroid that is  non-existing in the
conventional spherical distribution of the same orbitals.

Evidently, individual toroidal polarizations are,
individually, extremely unstable because the spherical distribution is recovered in
nanoseconds following the removal of the external magnetic field due to temperature related effects.
Nevertheless, when two toroidal polarizations are bonded together by opposing magnetic polarities
North-South-North-South- etc. as in Figure 3, spherical distributions are again
recovered in nanoseconds following the removal of the external magnetic field, but this
time such distribution occurs for the bounded pair as a whole.

The experimental detection of magnecules is rather difficult since it requires
analytic instruments and methods different than those currently used to detect molecules. Vice
versa,  analytic methods so effective to detect molecules generally reveals no magnecules, and this
explains their lack of detection since the discovery of molecules in the mid of the 19-th century.

An analytic equipment developed for molecules that is also effective for the detection of
gaseous (liquid) magnecules is given by a Gas (Liquid) Chromatographer Mass Spectrometer necessarily
equipped with InfraRed Detector for gases (GC-MS/IRD) or with UltraViolet Detector for liquids
(LC-MS/UVD).

Let us recall that large clusters (of the order of hundreds of amu or more) cannot be
constituted by molecules when without an IR signature for gases or a UV signature for liquids,
because that would require perfect spheridicity that is prohibited by nature for a large
number of constituents.



\begin{figure}
\begin{center}
\epsfxsize=4cm
\parbox{\epsfxsize}{\epsffile{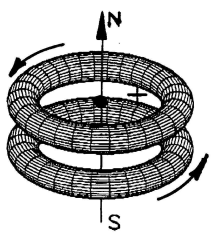}}
\end{center}
\caption{{\it A schematic view of a ''diatomic Santilli magnecule'' consisting of the bonding of two
atoms caused by the attractive force between opposing polarities North-South-North-South- etc. of
toroidal polarizations of at least some peripheral atomic electrons [9c,9d]. Note that, in reality,
the ''magnecular bond'' is rather complex since it is characterized by the attraction among
''three'' magnetic moments (those of the toroids, plus the intrinsic magnetic moments of the
electrons and of the nuclei), as well as the repulsive force among equal nuclear and electron
charges. Consequently, the figure depicts a condition of equilibrium between these opposing forces.
Note also the absence in magnecular bonds of considerations pertaining to the nature of the atoms
and the possible availability of valence electrons. Note finally the lack of limits in the number
of constituents in magnecules except limits set by instabilities due to collisions. Note that
Santilli's magnecules naturally realize the ''axial'' (but not the ''planar'') compatible spin
coupling of Figure 2 (see monograph [3l] for comprehensive studies). }}\label{Fig3}
\end{figure}


The detection of a magnecule requires its identification, firstly, with a peak in the MS that must
result to be unknown following the computer search among all known molecules and, secondly, that
peak must show no IR or UV signature at its amu value. The latter condition explains the need for a
GC-MS (or LC-MS) necessarily equipped with IRD (UVD). In fact, if the same species is
tested with an IRD (or UVD) disjoint from the MS, the IRD (UVD) is not generally focused on the
selected MS peak at its amu value, resulting in the detection of a variety of signatures of
conventional molecular species that, in reality, are the {\it constituents} of the
considered magnecule. Note that the lack of IR or UV signature also confirms the achievement of the
desired {\it bond weaker than the valence,} as needed to achievement full combustion (see, for
details, website [9c]).

As indicated in Section 2, the word ''valence'' is essentially a nomenclature due to the lack of
explicit and concrete identification of the ''attractive'' force necessary to produce a valance
bond (for Santilli-Shillady strong valence force as in Eq. (2.19), see Refs. [9a,9b,3l]). By
comparison, Santilli identified in the original proposal [9c] the {\it attractive character} of the
magnecular forces as well as its {\it numerical value,} that was confirmed by Kucherenko and
Aringazin [9e] as well as by others [3l].

The importance of the new species of magnecules for controlled nuclear fusions is established by an
inspection of Figure 3, where one can see that the toroidal polarizations of the peripheral orbitals
does indeed expose nuclei, as desired. The configuration clearly result to be preparatory
for the subsequent nuclear synthesis. Finally, the absence of IR signatures for gases or
UV signatures for liquids confirms that the bond occurs at low energy, as necessary for
controlled nuclear fusions. We therefore have the following:
\vskip0.30cm

{\it DEFINITION [9c,3k,3l,]: Santilli's magnecules are stable clusters consisting of individual
atoms ($H, C, O,$ etc.), dimers ($OH, CH,$ etc.) and ordinary molecules ($CO, H_2O,$ etc.)
bonded together by opposing magnetic polarities originating from toroidal polarizations of the
orbitals of atomic electrons.}
\vskip0.30cm

Numerous new substances with magnecular structures have been identified experimentally to date,
among which we indicate MagneGas$^{TM}$ [9d], MagneHydrogen$^{TM}$ [9h], $HHO^{TM}$ [9i], and others
under industrial development. Their primary features (for which large industrial investments have
been made) is the complete combustion without contaminant in the exhaust and cost
competitiveness over fossil fuels.

It is now customary in the field to denote a molecular bond with the symbol ''$-$'' and a magnecular
bond with the symbol ''$\times$.'' Consequently, the hydrogen molecule is represented with $H_2
= H-H$, while hydrogen magnecules are represented with the symbol $MH = H\times H\times ...$. A main
difference is that the only possible valence bond is $H_2$ (trivially, because the hydrogen atom
has only {\it one} electron), while there is no theoretical limit for the number of constituents
under magnecular bond except those set by collisions. In fact, a species of MagneHydrogen
$H_{14}$ having  {\it seven} times  the amu of $H_2$ has been detected in independent laboratories
[9h]. The latter measurements provide final confirmation on the existence of magnecules due to the
evident impossibility of any credible interpretation via valence.
\vskip0.50cm


\noindent {\bf 5. Proposed Industrial Realization of Controlled Intermediate Fusions via Hadronic
Reactors.} Without any claim of completeness or uniqueness, in this section we propose in the
necessary construction details a concrete {\it hadronic reactor} (patented and international patents
pending),that is, an equipment for the possible industrial utilization of new clean energies
produced by Intermediate Energy Controlled Nuclear Fusions (IECNF), or ''intermediate fusions'' (IF) for short,
via the engineering implementation and optimization of all seven Santilli's Laws on Controlled
Nuclear Fusions of Section 3. The application of the results to Low Energy Controlled Nuclear Fusions (LECNF),
or ''cold fusions' (CF)' for short, will be left to interested readers. High Energy Controlled Nuclear Fusions
(HECNF), or ''hot fusions'' (HF) for short, shall be ignored because outside realistic feasibility
based on current scientific knowledge and technological capabilities.

To begin, we use nature, rather than pre-existing research [1], for guiding lines. As established by
chemical analyses of air bubbles in amber, about one hundred millions years ago Earth's atmosphere
had about $40\%$ of nitrogen, while its current percentage is about double that value. Other chemical
analyses show that the increase of nitrogen in our atmosphere has been gradual. These data
establish {\it the existence in our atmosphere of a process for the  natural synthesis of
nitrogen from lighter elements.}

Among all possible origins of such a synthesis, the most probable is given by {\it lighting,}
because a serious scientific (that is, quantitative) explanation of {\it thunder} requires nuclear
syntheses. In fact, a numerical explanation of one thunder requires energy equivalent to hundreds
of tons of explosives that simply cannot be explained via conventional processes due to the very
small cylindrical volume of air affected by lightning combined with its extremely short duration of
the order of nanoseconds (serious scholars are suggested to do these calculations  to prevent
venturing nonscientific opinions).

By comparison, a relatively low rate of nitrogen syntheses provides indeed a numerical explanation
of thunder as well as its slow rate of increase in our atmosphere. Among all possible syntheses, the
most probable results to be the {\it synthesis of nitrogen from carbon and deuteron.} Needless to
say, numerous alternative fusions are also possible and some of them will be indicated below.

Consequently, by following nature, in this section we propose a specific and concrete
industrial realization of hadronic reactors that, by central conception, is based on
processes associated to lighting and thunder. Therefore, we have the following optimization of
Laws I-VII for the specific objective at hand:

\vskip0.30cm

{\bf Optimization of Law I:} For the preferred embodiment identified below, the implementation
and optimization of energy conservation can be achieved by controlling the temperature of the
chemical species selected for IECNF.

{\bf Optimization of Law II:} For simplicity, as well as in order to operate at the {\it lowest}
possible energies, in this section we shall select engineering realizations and optimization
applicable under the conservation of the angular momentum, with the understanding that the
restriction is not scientifically necessary for the conditions at hand in which the Poincar\'e
symmetry and special relativity are inapplicable, thus permitting a variety of additional hadronic
reactors (that would be suppressed by a nonscientific imposition of special relativity for
conditions it was not built for with evident damage to society) we plan to address in a subsequent
paper.
\vskip0.30cm



\begin{figure}
\begin{center}
\epsfxsize10cm
\parbox{\epsfxsize}{\epsffile{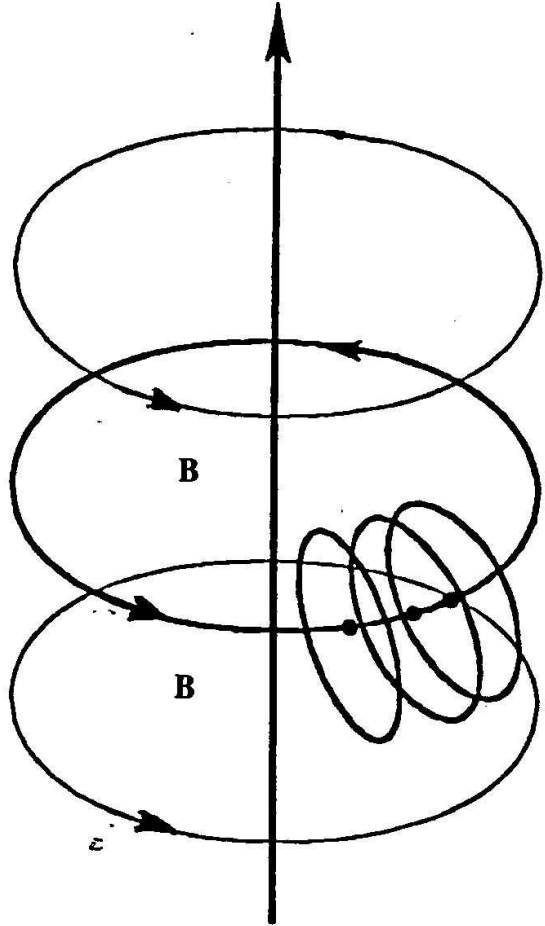}}
\end{center}
\caption{{\it A schematic view of the geometry of a DC electric arc represented by the vertical line,
with the associated magnetic field represented by horizontal circles, and the created magnecules
represented by circles perpendicular to the latter. This geometry has the following primary
implications: 1) Since the magnetic field $M$ is proportional to
$I / r$, one can see that at atomic distances $r = 10^{-8} cm$ from electric arcs with $ I = 10^3 A$
the magnetic field is of the order of $10^{11} G$, thus being amply sufficient to polarize atomic
orbitals [3l,9f,9g]; 2) Following said polarization, the geometry of electric
arcs is such to align automatically polarized atoms with opposing polarities
North-South-North-South-..., thus creating magnecular bonds automatically possessing the axial spin
couplings of Figure 2; and 3) For reasons not entirely understood [9c], electric arcs
compress magnecules toward their axis at the time of their initiation of shut off, thus assisting in
the realization of the trigger necessary for nuclear fusions.}}\label{Fig4}
\end{figure}


\vskip0.30cm
{\bf Optimization of Law III:} With reference to Figure 2, there are two types of engineering
implementation and optimization of the condition to have compatible spins, called in the
literature of hadronic mechanics [3i,3k] {\it planar and axial  compatible couplings.}
The engineering ''implementation'' of the planar
coupling can indeed be achieved (see the various proposals of Refs. [3i,3k,3l]), e.g., by subjecting
to opposing polarizations  two nuclear beams. However, the ''optimization'' of Law III definitely
suggests the adoption of the compatible axial coupling over the planar one for various reasons, such
as the fact that, for the case of planar coupling, the control of the polarization is lost at the
initiation of the fusion, with evident dispersal and loss of efficiency, while the
axial coupling can be controlled all the way to the completion of the fusion, with evidently
higher efficiency. Therefore, the preferred embodiment depicted below is based on the engineering
implementation and optimization of  Law III via compatible axial couplings.

\vskip0.30cm

{\bf Optimization of Law IV:} For the preferred embodiment of this section, the engineering
implementation and optimization of the minimal possible threshold energy is also
achievable via the control of the temperature and other features discussed below.
\vskip0.30cm

{\bf  Optimization of Law V:} The implementation of this law is achieved by
selecting nuclei in such a way that {\it the original as well as final nuclei are natural and
stable isotope.} The ''optimization'' of this law definitely favors {\it light, natural and stable
nuclei} over heavier ones for various reasons, e.g., the fact that the heavier the nuclei, the
bigger the possibilities for secondary radiations.

\vskip0.30cm

{\bf Optimization of Law VI:} As indicated in the preceding section, the hadronic
reactors proposed in this paper are based on the creation of a magnecular bond
prior to the nuclear fusion because this new bond automatically verifies Laws I,
II, III, IV, V and VI. However, the creation of the new species of magnecular is not elementary
because as studied by Aringazin [9f], the
polarization of electron orbitals to create magnecular bonds requires  magnetic fields so intense (of
the order of $10^{10} G$ or more) that cannot be provided by the most powerful  laboratory magnets.
The solution adopted by Santilli [9d] in the original proposal of the new chemical species of
magnecules is the use of  {\it  flowing a selected fluid through a submerged electric arc  so as to
continuously remove magnecules from the arc soon after their formation} (this is the so-called
''PlasmaArcFlow$^{TM}$ Process [9d] - US Patented and international patents pending). In fact,
the magnetic field surrounding electric arcs has indeed the intensity necessary for the toroidal
polarization of the orbitals. The continuous removal of the magnecules from the arc is then necesary
for control of the process.

\vskip0.30cm

{\bf Optimization of Law VII:} As it is well known in the new field of clean burning
fuels, magnecular bonds such as that of Figure 4 {\it
cannot} yield nuclear fusions, trivially, because the two nuclei have the same charge,
thus experiencing an intense Coulomb repulsion. Magnecular structures such as that of
Figure 3 essentially consist of a new statistical equilibrium among a variety of
electromagnetic forces. In order to convert a magnecular bond as that of Figure
3 into a nuclear fusion, there is the need of an external mechanism (the ''trigger'') that forces the
two nuclei at mutual distances of the order of $10^{-13}
cm$ (the ''hadronic horizon''), at which point the new strongly attractive
forces identified by
hadronic mechanics and chemistry take effect and, under the verification of all preceding laws,
the fusion is inevitable. The ''trigger'' adopted in this proposal is given by a combination of {\it
pressures} as well as {\it pulse DC arcs.}
\vskip0.30cm

The achievement of ''intermediate fusions'' of
industrial value requires the systematic production of energy in a reliable and
repetitive way without excessive service interruptions. The ''implementation'' of this requirement
eliminates liquids as feedstocks of hadronic reactors because of the short life of the electrodes
needed for the creation of magnecular bonds and other reasons. In fact, arcs within liquids
can only occur at very short distances proportional to the arc power, thus exposing the
electrodes to the large energy of the fusion, with their consequential rapid disintegration and
lack of industrial maturity. More generally, the ''optimization'' of the requirement here considered
requires the abandonment of a rather general tendency in the field [1], that of materializing
nuclear fusions inside the electrodes themselves. In fact, this approach prevents any possible
industrially viable engineering because of the extremely short life of the electrodes, let alone
their cost. To avoid these problem, the ''optimization'' here selected is that {\it ''intermediate
fusions'' are created by the arc itself and not by the electrodes.}
\vskip0.30cm



\begin{figure}
\begin{center}
\epsfxsize=18cm
\parbox{\epsfxsize}{\epsffile{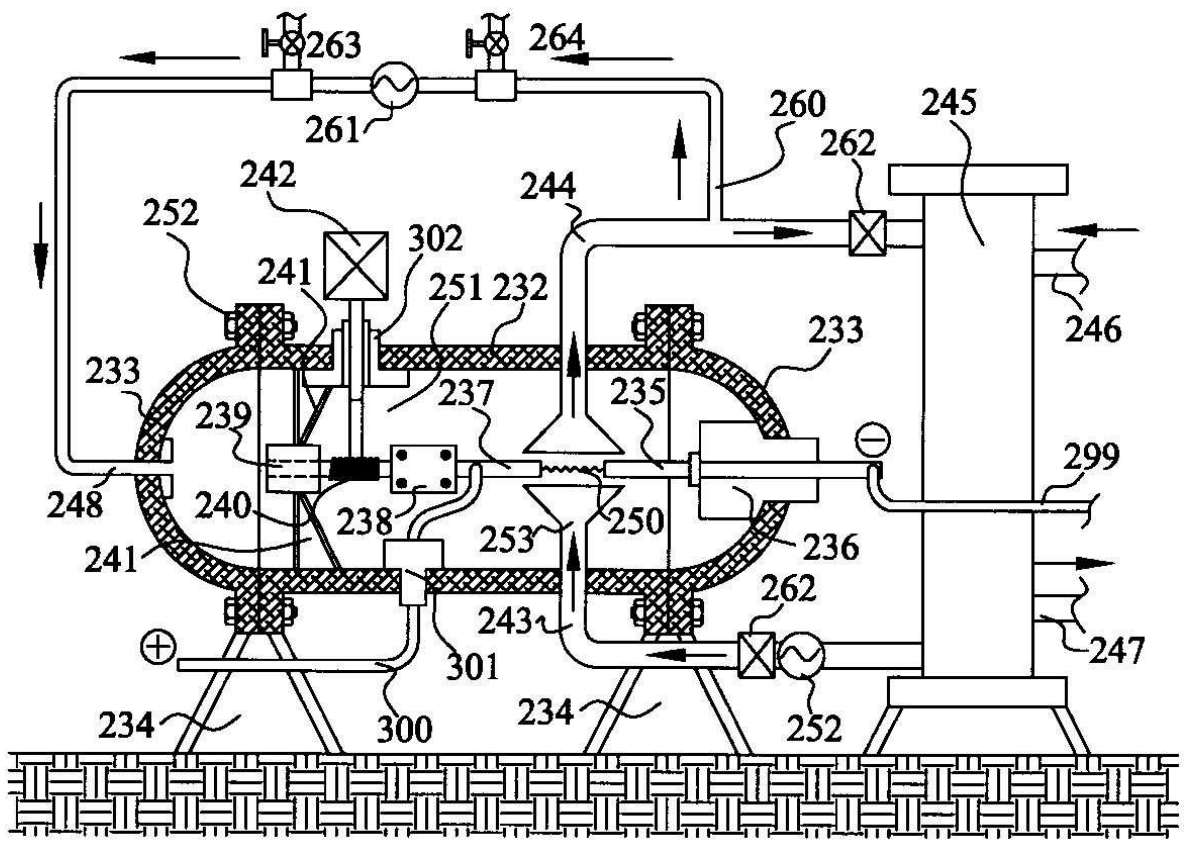}}
\end{center}
\caption{{\it A schematic view of the preferred embodiment for the industrial realization of the
proposed controlled intermediate fusion. }}\label{Fig5}
\end{figure}


By using the above optimization, a preferred embodiment of the hadronic reactors herein proposed
consists of a metal vessel capable of withstanding steady pressures up to $10,000 psi$ ($666 bars$)
as well as impulse pressures up to $100,000 psi$ ($6666 bars$) in which a 50 Kw DC electric arc of
steady and impulse nature is initiated, maintained and optimized via the automatic
controls of the Magnegas Technology [9d]. The vessel is filled up with a {\it gaseous} (rather than
liquid) feedstock as selected below and continuously recirculated through said arc. The control of
the energy output is done by controlling: 1) the value and frequency of the impulse pressure; 2) the
power and the frequency of the pulse DC arc; and 3) the flow of the fluid through the arc.

With respect to Figure 5, the proposed hadronic reactor (patented and international patents pending)
comprises: a metal vessel 232 with hemispherical heads 233 and fasteners 252 and bases
234 capable of withstanding a steady pressure of at least 10,000 psi (666 bars) and an impulse
pressure of at least b100,000 psi (6666 bars); a stationary, negatively charged, tungsten anode 235
that protrudes outside the hemispherical head  233 for connection via cable 299 to the negative
polarity of a steady or pulsing AC-DC converter with at least 50 Kw power (not shown in the
figure), said protrusion occurring  through insulating pressure resistant bushing 236 in phenolic
G10 or equivalent; an internally movable, positively charged tungsten cathode 237 connected via
cable 300 and insulating bushing 301  to the positive polarity of said outside power
source; said cathode 237 being connected via insulating phenolic G10 block 238 to a metal rod
equipped with rake 239 that is internally fastened to vessel 232 via brackets 240; said rake 239
being operated by a pignon 240 that is controlled by an outside servomotor 242 through
insulating pressure resistant bushing 302; vessel 232 being filled up with a gaseous
feedstock  251 that is recirculated through the electric arc 250 via blower 252 through
pipe 253; the gaseous feedstock is then sucked by pipe 244 for passage through heat
exchanger 245 for continuous recirculation through the arc 250 via blower 252 and pipe
253; the heat acquired by heat exchanger 245 being utilizes via an external fluid via
inlet 246 and outlet 247; the proposed hadronic recycler being completed by pipe 248 for
burst of pressure of the gaseous feedstock inside vessel 232 to realize the hadronic
trigger, said burst of pressure being realized by outlet 260 and impact blower 261,
the two check valves 262 protecting the primary blower 252 and the heat exchanger 245.

The operation of the proposed hadronic reactor is the following. Firstly, a high vacuum inside
vessel 232 is secured via valve 263. Subsequently, valve 263 is closed and the vessel is filled up
with the gaseous feedstock  251 via valve 264 up to the preset pressure of at least 10,000
psi (666 bars). At the achievement of the preset pressure, the automatic controls
activate the primary blower 252 and the continuous recirculation of the gaseous
feedstock through the arc is established. DC power is then automatically released to the
anode-cathode pair when the electrodes are at such a distance not to allow an arc for the pre-selected
gaseous feedstock and for the pre-selected pressure (open arc). Via the use of servomotor 242 acting
on pignon 240 that, in turn activates rake 239 solidly connected to cathode 237 via
insulating bushing 238, the automatic control move said cathode 237 toward the
stationary anode 235  until such a distance at which an electric arc of high current (e.g., 1,000 A)
within said gaseous feedstock is activated. This first phase serves to create magnecules. The
automatic controls then increase the gap between the electrodes to such a value for which the
variation of the voltage is within preset values (one of the twenty adjustable parameters of the
automatic controls of the Magnegas Technology [9d]), so as to maximize the travel of the arc within
the gaseous feedstock for an electric arc with present stability. Following a preset duration of
such high current arc, the automatic control active the high voltage impulse current as a partial
realization of the trigger. According to a pre-set frequency, the automatic control also activate the
impulse blower 261 to create burst of very high pressure inside vessel 232.a the trigger via a
combination of the following three means: 1) Impulse high voltage arcs; 2) Impulse high
pressures; and 3) the enhancement of both preceding contributions by the arc geometry (Figure 4).
It would be naive to assume that the above description is exhaustive, since numerous other features
are needed to render the above hadronic reactor industrially viable, but they are
omitted here for security reasons.

The desired Intermediate Energy Controlled Nuclear Fusions are of the
generic type
$$
TR +  N_1(A_1, Z_1,J_1^{p_1}) + N_2(A_2, Z_2, J_2^{p_2}) \; \; \; \rightarrow \; \; \; N_3(A_3,
Z_3, J_3^{p_3})  +  Heat,
\eqno(5.1)
$$
where: TR is the trigger; $A$ is the number of nucleons; $Z$ is the number of protons; $J$ is the
angular momentum; $p$ is the parity;
$A_1 + A_2 =  A3,    Z_1 + Z_2 = Z_3, J_1 + J_2 = J_3, p_1 = p_2 = p_3$; and, by central
assumption, {\it the original and final nuclei are light, natural, and stable isotope.}

To illustrate how restrictive Laws I-VII are, it is important to show that {\it the use of the
deuteron gas for the synthesis of the helium is not recommendable for the proposed hadronic
reactor.} In fact, Eq. (5.1) becomes in this case
$$
TR  +   H(2,1, 1^+) + H(2,1, 1^+) \; \; \; \not \rightarrow \; \; \; He(4,2, 0^+) + heat,
\eqno(5.2)
$$
that violates Law III on compatible spin coupling, as well as Law II on angular momentum
conservation.

A possible hadronic reactor for the synthesis of the helium verifying Laws II and III according to
the synthesis
$$
TR  +   H(2,1, 1^+) + H(2,1, -1^+) \; \; \;  \rightarrow \; \; \; He(4,2, 0^+) + heat,
\eqno(5.3)
$$
would  be dramatically different than that herein considered, and has been suggested
elsewhere [3k,3l]. Therefore, the synthesis of the helium is ignored hereon.

A more promising ''intermediate fusion'' is that of synthesizing a stable isotope of the Lithium from
a 50-50 mixture of deuteron and helium according to the following realization of Eq. (5.1)
$$
TR + H(2, 1, 1^+) + He(4, 2, 0^+) \; \; \; \rightarrow \; \; \; Li(6, 3, 1^+) + heat
\eqno(5.4)
$$
that verifies all seven laws of CNF.

A preferred use of the proposed hadronic reactor is that for the synthesis of nitrogen from carbon
and deuteron indicated earlier according to the fusion process
$$
Tr + C(12, 6, 0^+) + H(2, 1, 1^+) \; \; \; \rightarrow \; \; \; N(14, 7, 1+) + Heat,
]eqno(5.5)
$$
that verifies all seven laws of controlled nuclear fusions. The above fusion can be tested by using
tungsten electrodes and filling up the hadronic reactor with a 50-50 mixture of carbon dioxide and
deuteron gas. In this case, the electric arc decomposes $CO_2$ into carbon and oxygen, thus
rendering the carbon available flr fusion (5.5). The resulting oxygen is also expected to have
''intermediate fusions'' via the reaction
$$
TR + O(16, 8, 0^+) + H(2, 1, 1^+) \; \; \; \rightarrow \; \; \;  F(18, 9, 1+),
\eqno((5.6)
$$
that also verifies all CNF laws, with the exception that $F(18, 9, 1+)$ is not a stable isotope.
Nevertheless, it decays in about $109$ minutes into the oxygen via an electron capture or a beta
plus  decay, thus being acceptable on environmental grounds (since the beta are easily trapped by
the heavy steel of the reactor vessel). Another alternative is the use of carbon electrodes and then
filling up the hadronic reactor with only deuteron gas. In this case the electrodes will consume
since they provide the carbon needed for synthesis (5.5), although their cost is minimal and fast
means of their replacement are possible for minimal service [9d], to as to maintain industrial
maturity.

Numerous other gaseous feedstock are possible for the proposed hadronic reactor. Their systematic
study is left to interested readers for brevity.

The expected energy output of the nitrogen synthesis is significant. In fact, we have the energy
release per synthesis
$$
\Delta E = [C(12, 6, 0^+) + H(2, 1, 1^+)]  - N(14, 7, 1+) =
  14.850 MeV / c^2.
\eqno(5.7)
$$
By remember that $1 MeV = 1.6021\times 10^{-13}$ joule and
that in one mole we have  $6.022 \times 10^{23}$ atoms (Avogadro number), the
extremely low efficiency of one over $10^{7}$ atoms per mole per minute of said 50-50 mixture of
carbon dioxide and deuteron gas, would yield the energy release
$$
( 14.8 \times  10^6 \times 1.6 \times 10^{-19} Joule ) \times (  6 \times 10^{23} ) \times
( 10^{-7} \; \; reaction / min\; \; per \; \; mole ]  =  1.4 \times 10^6 joule / min,
\eqno(5.8)
$$
namely, an energy outputm if confirmed, would have  full industrial significance since the energy
input ($50 Kw$) is essentially ignorable with respect to the above energy output on a per minute
basis.

To understand the engineering optimization of the proposed hadronic reactor it is important to
indicate other possibilities verifying all seven CNF laws, but they are not
recommended because of industrial insufficiencies. Suppose that the reactor is filled up with
hydrogen, and that the electrodes are made up of Palladium $106$ or $108$. In this case we expect
the following fusions {\it at the palladium cathode}
$$
TR + Pd(106, 46, 0^+) + H(1, 1, 1^+) \; \; \; \rightarrow \; \; \; Ag(107, 47, {1\over 2}^+),
\eqno(5.9a)
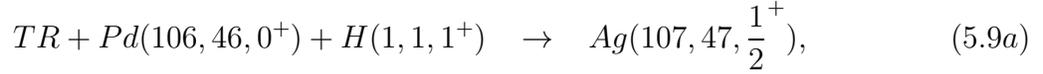
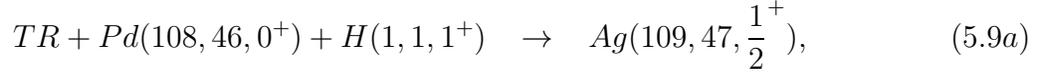
$$
$$
TR + Pd(108, 46, 0^+) + H(1, 1, 1^+) \; \; \; \rightarrow \; \; \; Ag(109, 47, {1\over 2}^+),
\eqno(5.9a)
$$
which fusions do verify all CNF laws. Nevertheless, the preceding fusions would imply the rapid
disintegration of the electrodes, with consequential lack of industrial relevance. This illustrates
the need for an embodiment to have a sufficiently long life prior to service as a necessary
condition for industrial maturity.

 A few comments are now in order. Firstly, {\it we stress the impossibility for the proposed
hadronic reactor to produce energy of explosive character,}  because  synthesis (5.5) occur
along the arc, thus displacing the gaseous feedstock away from the arc at their occurrence, with
consequential halting of all effects. This is the very reason why the patented PlasmaArcFlow
Technology is mandatory to reach industrially meaningful results.

Secondly, note the {\it the impossibility for fusion (5.5)
to produce any harmful radiation,} evidently because either the nitrogen is synthesized or not,
while the emission of neutrons is impossible because the available energies are dramatically
insufficient for the fission of any available nucleus, while possible proton and electron radiations
are easily trapped by the heavy metal vessel due to their charges.

Thirdly, one should
note that the energy output is easily controllable in the proposed hadronic reactor in a variety
of way, including the control via the values and impulse frequency of pressure and DC power as well
as the control of the PlasmaArcFlow.

Despite these intrinsic safety features, all energy productions imply risks, and this is the case
also for the proposed hadronic reactor. In fact, the latter can only operate at high pressures, thus
requiring safety precautions for any operation, essentially given by operation under ground with
heavy steel reinforced cement protections due to known risks connected to high pressure, and the sole
possible operation via completely automatic remote controls. This illustrates the need of the
proposal to use the already developed automatic and remote controls of the Magnegas Technology.

We should also indicated that the proposed hadronic reactor is based on preliminary experimental
evidence of the MagneGas Reactors of Ref. [9d] in regard to the production of anomalous heat (that
is, heat that cannot be entirely accounted with conventional thermochemical reactions), as well as
anomalous content of nitrogen in Magnegas. Nevertheless, these indications should be taken with care
due to the need of systematic measurements for their independent verification not conducted until
now.

We would like also to confirm that hadronic mechanics, hadronic chemistry and isorelativity do
indeed predict that, under the above realization and optimization of all seven CNF laws via the
proposed hadronic reactor, the  synthesis of the nitrogen from carbon and deuteron is
inevitable following the triggering though the hadronic horizon. However, we should also indicate
that the excessive number of unknowns due to the novelty of the research prohibit the prediction of
specific numerical values.
Their numerical values of pressures, DC power and PlasmaArcFlow have been suggested above on
semi-empirical grounds based on their maximal possible engineering realization. Therefore, no claim
of actual existence of the proposed ''intermediate fusion'' of nitrogen from carbon and deuteron can
be voiced prior to the actual construction and successful test of the proposed hadronic reactor.

In closing, it is appropriate to recall that both the ''cold'' and the ''hot'' fusions have produced
no industrially value result to date following large investments over a protracted period of time.
While research along these lines should evidently continue, pressing societal needs caused by ever
increasing cataclysmic climactic events requires serious research and investments on {\it new}
alternative, of which the ''intermediate fusion'' proposed in this paper is merely one among
other possibilities. In the final analysis, readers should remember that the well being of their
children is at stake.

\vskip0.50cm

\noindent {\bf Legal Note.}

\noindent The hadronic reactor proposed in this paper is protected by U.S. Patent numbers
6,926,872; 6,673,322; 6,663,752; 6,540,966; and 6,183,604; issued to R. M. Santilli and assigned to
private corporations plus a number of additional international patents are pending. This is to
acknowledge that, according to international patent laws, any  researcher is completely free to
conduct any and all desired {\it research} on the proposed  hadronic reactor without any need for
prior authorization from the patent holder, under the condition that this paper and the quoted
patents are listed in any possible publication. Prior authorization by the owners of the
intellectual rights is requested by law only in the event of sales or other uses of the patented
technology implying an {\it income.}

\end{document}